\documentclass[aps,prl,twocolumn,showpacs]{revtex4}

\usepackage{amsmath,amssymb,amsfonts}

\usepackage{inputenc}

\usepackage{graphicx}

\newcommand{\be}{\begin{equation}}
\newcommand{\ee}{\end{equation}}

\begin{document}

\title{Predicted and Verified Deviation from Zipf's Law in Growing Social Networks}

\author{Qunzhi Zhang and Didier Sornette}
\affiliation{ETH Zurich,
 Department of Management, Technology and Economics, Kreuzplatz 5, 
CH-8092 Zurich, Switzerland}

\begin{abstract}
Zipf's power law is a general empirical regularity found in many natural and social systems. A
recently developed theory predicts that Zipf's law 
corresponds to systems that are growing according to a maximally sustainable path
in the presence of random proportional growth, stochastic birth and death processes.
We report a detailed empirical analysis of a burgeoning network of social groups,
in which all ingredients needed for Zipf's law to apply are verifiable and verified. 
We estimate empirically the average growth $r$ and its standard deviation $\sigma$
as well as the death rate $h$ and predict without adjustable parameters the exponent $\mu$
of the power law distribution $P(s)$ of the group sizes $s$. The predicted
value $\mu = 0.75 \pm 0.05$ is
in excellent agreement with maximum likelihood estimations. According to theory,
the deviation of $P(s)$ from Zipf's law (i.e., $\mu < 1$) 
constitutes a direct statistical quantitative signature of the 
overall non-stationary growth of the social universe. 
\end{abstract}
\date{\today}
\maketitle

Power law distributions,
\begin{equation}
p(s) \sim 1 / s^{1+\mu}~,
\label{wbtwr}
\end{equation}
are ubiquitous characteristics
of many natural and social systems. The function $p(s)$ is the density
associated with the probability $P(s)= {\rm Pr}\{S>s\}$ that 
the value $S$ of some stochastic variable, usually a size or
frequency,  is greater than $s$. Among power law distributions,
Zipf's law states that $\mu=1$, i.e., $P(s)\sim
s^{-1}$ for large $s$.  Zipf's law has been reported for many systems \cite{dsornette}, including 
word frequencies \cite{Zipf49}, firm sizes \cite{Axtell01},
city sizes \cite{Gabaix99}, connections between Web pages \cite{Kongetal08} and 
between open source software packages \cite{tmaillart}, Internet traffic characteristics \cite{Huberman1},
abundance of expressed genes in yeast, nematodes and human tissues \cite{FurusawaKaneko03} and so on.
The apparent ubiquity and universality of Zipf's law has triggered numerous efforts
to explain its validity. Deviations from Zipf's law
provide also important informative insights in the dynamics of the corresponding systems.

Since H. Simon's pioneering work \cite{Simon55,Simon60,IS77}, a crucial
ingredient in the generating mechanism of Zipf's law is understood to be Gibrat's rule
of proportional growth \cite{Gibrat31}, more recently rediscovered
under the name of ``preferential attachment'' in the context of networks \cite{AlbertBaraAlbert99}.
Expressed in continuous time in terms of the size $S(t)$ of a firm, a city or, 
more generally, a social group, Gibrat's rule corresponds
to the geometric Brownian motion
\be
dS(t) = S(t) \left( r \, dt + \sigma \,  dW(t)\right)~,
\label{thynj4u6j}
\ee
where the stochastic growth rate $r+ \sigma dW/dt$ is decomposed
into its average $r$ and its fluctuation part with amplitude determined by
the standard deviation $\sigma$, while $W(t)$ is a standard Wiener process.
Gibrat's rule alone cannot produce (\ref{wbtwr}), since the solution of equation (\ref{thynj4u6j}) 
has a log-normal distribution. Simon and many other authors invoked an addition ingredient,
corresponding to various modifications of the multiplicative process when $S(t)$ becomes small.
Then, under very general conditions, the distribution of $S$ becomes a power law, with
an exponent $\mu$ that is a function of the distribution of the multiplicative factors \cite{Kesten73,Sornette98,Gabaix2009}.

The fact that the exponent $\mu$ is often found close to $1$ requires another 
crucial ingredient.  One particularly intriguing proposition is that Zipf's law
corresponds to systems that are growing according to a maximally sustainable path \cite{dsornette2}.
In other words, when Zipf's law holds, the set of stochastically
growing entities $\{S_i(t), i=1, 2, ..., n, ..\}$ is delicately poised at a dynamical critical growth point. Within 
a general framework in which (i) entities are born at random times, (ii) grow
stochastically according to (\ref{thynj4u6j}), and (iii) can disappear or die according
to various stochastic processes with some hazard rate $h$, 
the explicit calculation of the exponent $\mu$ confirms the above optimal growth condition
associated with Zipf's law ($\mu=1$) \cite{dsornette2}.

Here, we present an empirical test of the optimal growth condition for Zipf's law 
by testing the formula for exponent $\mu$ (see below) on a unique
database obtained from a Web platform of collaborative social projects (Amazee.com).
In this dataset, we verify empirically that proportional growth holds, we measure the parameters
$r, \sigma$, $h$ and the exponent $\mu_{\rm exp}$ of the power law distribution of project sizes.
We show that the theory leading to the maximum sustainable growth principle
explains remarkably well the empirical value $\mu_{\rm exp}$, with no adjustable parameters.

The theory  is based on the following assumptions \cite{dsornette,dsornette2}.
Consider a population of social groups (firms, cities, projects, and so on), which can take different forms and can be applied
in many different contexts. 
\begin{enumerate}
\item There is a flow of group entries, i.e., a sequence of births of new groups. The times
$\{t_1 < t_2 < ...  < t_i< ...\}$ of entries of new groups follow
a Poisson process with constant intensity (generalizations 
to a vast class of non-Poisson processes do not modify the key result \cite{dsornette,dsornette2}).
\item At time $t_i$, $i \in {\mathbb N}$, the initial size of the new entrant 
group $i$ is a random variable $s_{0,i}$. The sequence
$\left\{s_{0,i}\right\}_{i \in {\cal N}}$ is the result of independent and identically distributed random 
draws from a common random variable $\tilde s_0$. All the draws are independent of the entry dates of the
groups.
\item Gibrat's rule of proportional growth holds. This means  that, in the continuous time limit, the size $S_i(t)$ of the $i^{th}$
group at time $t \ge t_i$, conditional on its initial size $s_0^i$, is solution to the stochastic differential equation
(\ref{thynj4u6j}), where the drift $r$ and the volatility $\sigma$ are the same for all groups but 
the Wiener process $W_i(t)$ is specific to each project $i$.
\item Groups can exit (disappear) at random, with constant hazard rate 
$h \ge 0$, which is independent of the size and age of the group.
\end{enumerate}
Under these conditions,  the central result of \cite{dsornette2} reads as follows. Defining 
\be
\label{eq:m}
\mu := \frac{1}{2} \left[\left(1 - 2 \cdot \frac{r}{\sigma^2}\right) + 
\sqrt{\left(1 - 2 \cdot \frac{r}{\sigma^2}\right)^2 + 8 \cdot \frac{h}{\sigma^2}} \right]~,
\ee
provided that ${\rm E}\left[\tilde s_0^\mu\right] < \infty$, and for times larger than 
\be
t_{\rm transient} = \left[ \left(r - \frac{\sigma^2}{2} \right)^2 + 2 \sigma^2 h\right]^{-1/2}~,
\label{yjyjn}
\ee
 the average distribution of project's sizes follows an asymptotic power law 
 with tail index $\mu$ given by (\ref{eq:m}),  
 in the following sense: the average number of projects with size larger than $s$ is 
 proportional to $s ^{-\mu}$ as $s \to \infty$. As a corollary, the exponent $\mu$ of the distribution of sizes takes the value $1$
 corresponding to Zipf's law, if and only if $r = h$.
 
 In order to understand the corollary, notice that 
$r - h$ represents the average growth rate of an incumbent group. 
Indeed, considering a group present at time $t$, during the next instant $dt$, it will either exit 
with probability $h \cdot dt$ (and therefore its size
declines by a factor $-100\%$) or grow at an average rate equal to $r \cdot dt$, with probability 
$(1- h \cdot dt)$. The coefficient $r$ is therefore the conditional growth rate
of projects, conditioned on not having died yet. Then, the unconditional expected growth rate over the 
small time increment $dt$ of an incumbent group is $(r-h) \cdot dt + O\left((dt^2\right)$. 
The statistically stationary regime,
in the presence of a stationary population of group forming individuals,
 corresponds to condition $r = h$.
Malevergne et al. \cite{dsornette2} showed that this condition
can be easily generalized to the case where the population of group forming individuals 
grows itself with some exponential rate, as is the minimal viable group size  \cite{dsornette2}.
Then, this condition translates into that for the maximum 
sustainable growth of the universe of groups, as mentioned above. 

Our strategy is to find an empirical dataset in which (i) all ingredients of the theory
can be verified explicitly, (ii) all parameters $r, \sigma$ and $h$ can be measured directly
and (iii) the empirical distribution of group sizes can be compared with to prediction 
(\ref{wbtwr}) with (\ref{eq:m}). We have found such a database, with 
Amazee.com, which is a Web-based platform of collaboration.
Using Amazee's Web-platform, 
anyone with an idea for a collaborative project can sign in and use 
the website to gather followers, who will together help the project
owner to accomplish the project. An Amazee project can be of any type of
activities, such as arts and culture, environment and nature, politics
and beliefs, science and innovation, social and philanthropic, sports
and leisure, and so on. Most of the projects are public, for instance, ``build
a strong community of Internet entrepreneurs in Switzerland to exchange
information and have fun'' (Web Monday Zurich), ``connect all women
working in the Swiss ICT industry'' (Tech Girls Switzerland), ``to
provide fresh running water to each home in the small African village of
Dixie'' (Water for Dixie), and so on. Amazee.com provides a set of
features covering the entire lifetime of a typical project, such as
project planning, participants recruiting, fund raising, events and
meetings hosting, communication, files archiving, and so on. Users join
Amazee.com by either creating a new project, or participating in projects
created by others. The Amazee data we analyze contains the complete
recording in time of the activities of all
users creating and joining all the projects in existence from February 2008
till May 2010.  

Projects can be seen as proxies of many naturally occurring entities,
such as social groups, firms, cities, investment vehicles, and so on, each driven by 
some goal, competition, and interaction within social networks. 
The detailed knowledge of the activity of the participants of all
projects provides a remarkable opportunity to dissect and understand
the dynamics of such systems. In the present study, we restrict
our attention to the simplest measure of size, namely the 
number $S_i$ of members of project $i$.

\begin{table}
\caption{Descriptive statistics of the sizes of Amazee's projects at different times, showing
that most projects have a size of just a few individuals while a few projects have 
hundreds to more than one thousand members. Dates are in format day/month/year.
}
\centering
\begin{tabular}{|l|l|l|l|l|}
\hline
 & 07.08.2008 & 08.02.2009 & 07.08.2009 & 08.05.2010\\
\hline
Number of projects & 436 & 1165 & 1562 & 1829 \\
\hline
Mean size& 5 & 10 & 9 & 8 \\
\hline
Minimum size & 1 & 1 &  1 & 1 \\
\hline
Maximum size & 83 & 1110 & 1114 & 1120 \\
\hline
Median size & 2 & 2 & 2 & 1 \\
\hline
\end{tabular}
\label{table:descriptive}
\end{table}

Amazee's platform started on February, 2008.
We analyze four snapshots of the database, on 7 August 2008, on 8 February 2009,
on 7 August 2009 and on 8 March 2010. The first snapshot
is six months after the birth of the operations on Amazee.com.
With the parameter values for $r, \sigma$ and $h$ determined below, formula (\ref{yjyjn})
predicts a transient of 50-400 days. Therefore, except
for the first snapshot, we should observe a reasonable convergence to the
expected power law distribution.

\begin{figure}
\centering
\includegraphics[width=0.5\textwidth,clip]{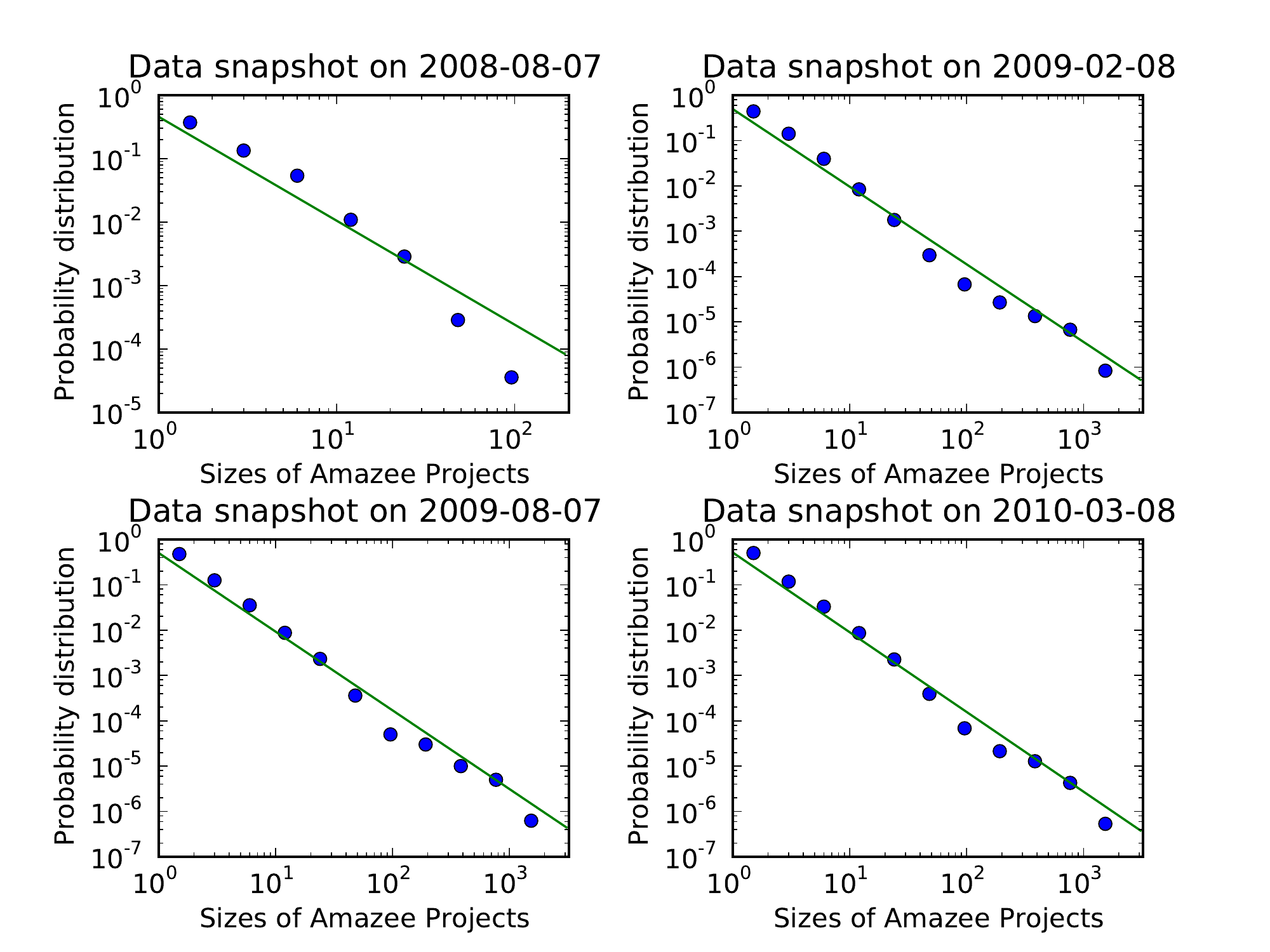}
\caption{Blue dots: Probability distributions of Amazee project sizes measured for four snapshots on  7 August 2008, on 8 February 2009,
on 7 August 2009 and on 8 March 2010. The maximum likelihood fits are shown by the green lines, with exponents $\mu$ 
respectively equal to $0.64$, $0.71$, $0.74$, $0.76$.}
\label{fig:pdf}
\end{figure}

Table \ref{table:descriptive} and Fig~\ref{fig:pdf} confirm
that the distributions of project sizes obtained for these
four snapshots are power laws (\ref{wbtwr}) ($p>0.05$ of a
Kolmogorov-Smirnov (K-S) test  for the
three last snapshots), with some significant
deviation only for the earliest snapshot. This deviation can be interpreted 
as only a partial convergence to the stationary growth regime, confirmed
by the much smaller maximum size observed in the first snapshot.
K-S tests on the same data for goodness of fit using competing distributions
such as the exponential distribution and the log normal distribution
yield $p$-values less than $0.01$, confirming the power law as
the best model.

\begin{figure}[!htp]
    \centering
    \includegraphics[width=0.5\textwidth]{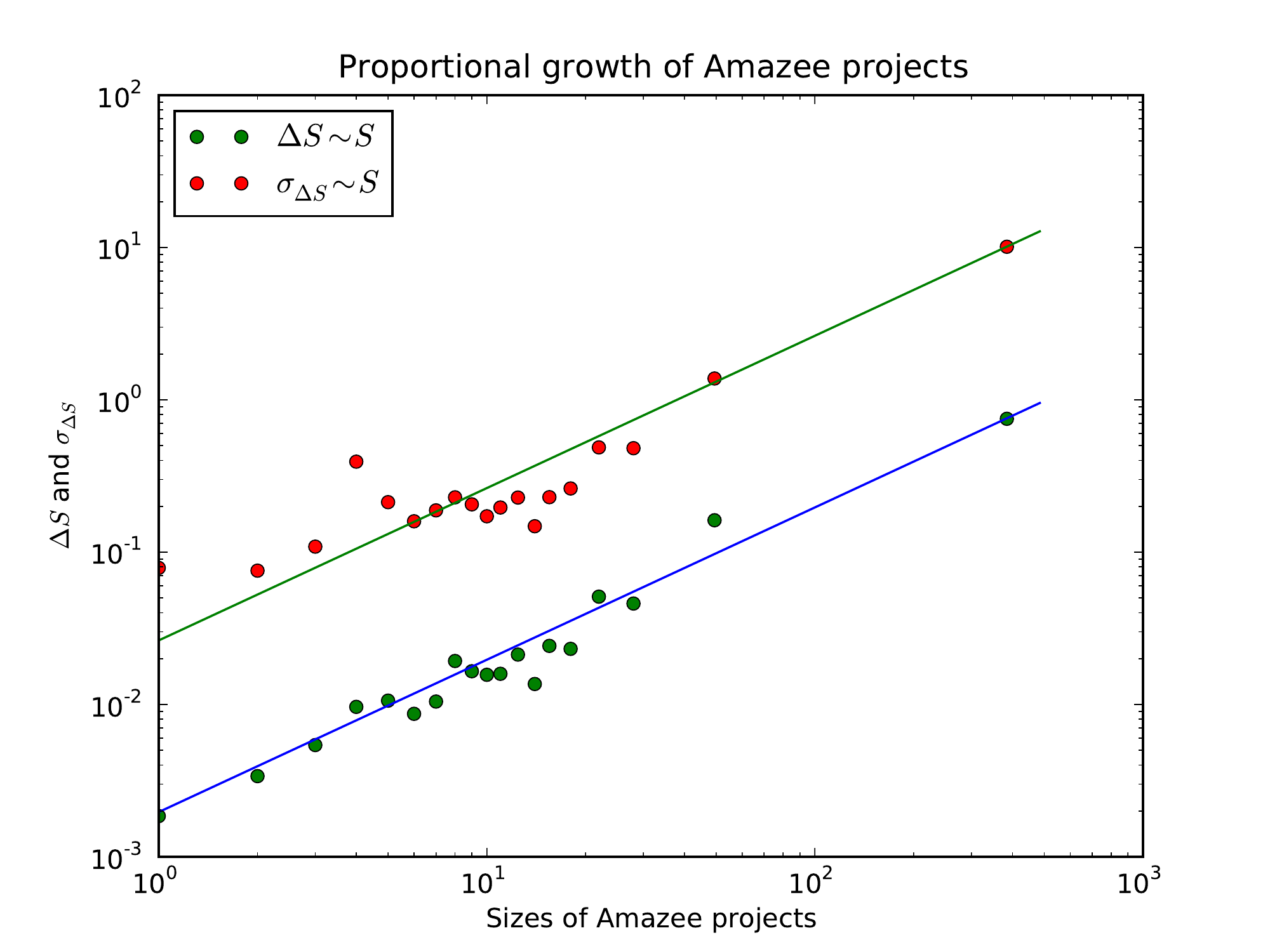}
    \caption{Test of Gibrat's law for the proportional growth of Amazee project sizes until 8 March 2010. 
    The slopes of the fitted straight lines are exactly $1$.}
    \label{fig:pgrowth}
\end{figure}

Because the numbers of project members are integers, the exponents
$\mu$ corresponding to the empirical distributions
shown in Fig~\ref{fig:pdf} are estimated using the maximum likelihood method (ML)
with the normalized discrete version
of (\ref{wbtwr}), $p(s)=\frac{s^{-(1+\mu)}}{\zeta(1+\mu)}$, where $\zeta(x)$ is the
Riemann zeta function: $\zeta(x)=\sum_{s=1}^\infty s^{-x}$.  The 
exponents are found around $0.7$, with confidence intervals clearly
excluding $1$. We check the robustness of this conclusion by estimating
the exponents $\mu$ for the four snapshots as a function of a lower threshold
above which the MLE is performed. For the three
last snapshots, we find stable estimations, with 
the $95\%$ confidence intervals excluding the value $1$.
We can thus conclude that Zipf's law is rejected for this dataset.   

\begin{table}[ht]
\caption{For each of the four snapshots of the amazee database, we report the parameters $r, \sigma$
and $h$ as explained in the text. Reporting these parameters in 
expression (\ref{eq:m}) yields the predicted exponents $\mu$, which is compared with the empirical
exponents $\mu$ estimated by maximum likelihood (MLE). For each cell value, 
the 95\% confidence interval is obtained by bootstrapping over 10'000 realizations.
Dates are in format day/month/year.}
\centering
\begin{tabular}{|c|p{2.0cm}|p{2.0cm}|p{2.0cm}|p{2.0cm}|}
\hline
Date & 07.08.2008 & 08.02.2009 & 07.08.2009 & 08.03.2010\\
\hline
$r$ & 0.11\newline [0.074, 0.20] & 0.031\newline [0.027, 0.036] & 0.027\newline [0.024, 0.031] & 0.019\newline [0.017, 0.021] \\
\hline
$\sigma$ & 0.30\newline [0.23, 0.41] & 0.18\newline [0.16, 0.20] & 0.18\newline [0.16, 0.20] & 0.19\newline [0.15, 0.24] \\
\hline
$h$ & 0.096\newline [0.065, 0.17] & 0.021\newline [0.019, 0.025] &  0.017\newline [0.015, 0.019] & 0.011\newline [0.0099, 0.012] \\
\hline
$\mu$ (MLE) & 0.64\newline [0.58, 0.70] & 0.71\newline [0.67, 0.76] & 0.73\newline [0.69, 0.78] & 0.76\newline [0.72, 0.80] \\
\hline
$\mu$ (TH)) & 0.89\newline [0.78, 1.05] & 0.78\newline [0.74, 0.81] & 0.73\newline [0.70, 0.75] & 0.75\newline [0.71, 0.79] \\
\hline
\end{tabular}
\label{table:predictedmu}
\end{table}

We now test formula (\ref{eq:m}). For this, we test if model (\ref{thynj4u6j}) holds and
proceed to estimating the parameters $r, \sigma$ and $h$. 
The proportional growth model posits that, for sufficiently
small time intervals $\Delta t$, the mean ${\rm E}[\Delta S]$ 
and the standard deviation $\sigma_{\Delta S}$ of the 
increment of the size $S$ of a given project
should both be proportional to $S$. To test this proposition, 
all the Amazee projects are pooled together in 100 size intervals
over all four snapshots. For each of the 100 size intervals, Figure~\ref{fig:pgrowth} 
plots the average daily increase of project sizes (${\rm E}[\Delta S]$) and its standard deviation
$\sigma_{\Delta S}$ as a function of $S$. 
Linear regressions give very high
$R^2$'s, larger than $0.995$, confirming that Gibrat's law holds.
The parameters $r$ and $\sigma$ are estimated
as the mean and standard deviations of the set of daily growth rates, and are 
reported in Table \ref{table:predictedmu}. 
Note that $\sigma_{\Delta S}$ is much larger than $\Delta S$, i.e., the stochastic
component of the proportional growth clearly dominates (an essential condition
for a power law to emerge in the model \cite{dsornette}). 

Next, we find that the rate of birth of new projects on amazee.com is approximately described by a Poisson process,
such that the probability that $n$ projects are born in a given day is given by
\begin{equation}\label{eq:births}
Pr\{n\}=\frac{\lambda ^ n}{n!}e^{-\lambda}~,
\end{equation}
where $\lambda \approx 2.4$ is the mean number of new born projects per day.  

Many projects eventually stop growing, when they have reached their goals or in the presence of operational problems.
We qualify a project as ``dead'' at some time $t_d$, if it has not added any new member for the 90 days
following $t_d$. If born at some time $t_b$, its lifetime $\ell$ is then calculated as $\ell := t_d-t_b$.
For projects with lifetimes of 12 days or larger, we find that the distribution of project lifetimes $\ell$ is very 
well approximated by the exponential law
\begin{equation}
\label{eq:deaths}
Pr\{\ell \geq T\}=e^{-h T}~,
\end{equation}
where $h$ is the death hazard rate, whose maximum likelihood estimations
are reported in Table \ref{table:predictedmu} for the four snapshots
of Amazee's database. A Kolmogorov-Smirnov test applied to (\ref{eq:deaths}) gives a p-value 
(estimated by bootstrap) of 93.7\%, 
confirming the exponential model (\ref{eq:deaths}).

Using the empirically determined values of $r, \sigma$ and $h$, we are now in position to test
the theoretical prediction (\ref{eq:m}) for the exponents $\mu$ 
of the proportional growth model in the presence of stochastic birth and death process.
As shown in Table~\ref{table:predictedmu}, except for the first snapshot
for which transient effects are present (as discussed before), the agreement is excellent,
with no adjustable parameters!

The detailed empirical analysis of the burgeoning social networks on Amazee has provided a unique
set-up to test predicted deviations from Zipf's law in a system in which all ingredients needed 
for Zipf's law to apply are verifiable and verified. The deviation from Zipf's law, namely
that the exponent $\mu$ is smaller than $1$, results from the fact that the average 
growth rate $r$ of Amazee projects is higher than their death rate $h$. Hence, 
the deviation from Zipf's law is a remarkable statistical signature of the 
overall non-stationary growth of the Amazee universe. 

After their time of fame and fashion, power law distributions have been sometimes decried as too general, perhaps
too universal to really provide useful insights. Here, we have provided an example in which 
the value of the exponent, and in particular its size less than $1$ is a direct fingerprint of 
the overall growth of a social system, under the combined actions of multiplicative noise, birth and death processes.
Given the generality of these ingredients, the prediction of the power law exponents provides new understandings of power law distributions, which will be insightful to many natural, economic and social systems. 

{\bf Acknowledgement}: We are grateful to Gregory Gerhardt, Thomas Maillart, 
Yannick Malevergne and Renaud Richardet for many stimulating discussions.
This work has been partially supported by the Swiss Federal Innovation Promotion Agency under grant
CTI 10442.2 PFES-ES entitled  ``Prediction and visualization of social dynamics''.

\bibliography{amazee}

\begin{thebibliography}{17}
\expandafter\ifx\csname natexlab\endcsname\relax\def\natexlab#1{#1}\fi
\expandafter\ifx\csname bibnamefont\endcsname\relax
  \def\bibnamefont#1{#1}\fi
\expandafter\ifx\csname bibfnamefont\endcsname\relax
  \def\bibfnamefont#1{#1}\fi
\expandafter\ifx\csname citenamefont\endcsname\relax
  \def\citenamefont#1{#1}\fi
\expandafter\ifx\csname url\endcsname\relax
  \def\url#1{\texttt{#1}}\fi
\expandafter\ifx\csname urlprefix\endcsname\relax\def\urlprefix{URL }\fi
\providecommand{\bibinfo}[2]{#2}
\providecommand{\eprint}[2][]{\url{#2}}

\bibitem[{\citenamefont{Saichev et~al.}(2009)\citenamefont{Saichev, Malevergne,
  and Sornette}}]{dsornette}
\bibinfo{author}{\bibfnamefont{A.}~\bibnamefont{Saichev}},
  \bibinfo{author}{\bibfnamefont{Y.}~\bibnamefont{Malevergne}},
  \bibnamefont{and} \bibinfo{author}{\bibfnamefont{D.}~\bibnamefont{Sornette}},
  \emph{\bibinfo{title}{Theory of Zipf's Law and Beyond, Lecture Notes in
  Economics and Mathematical Systems}}, vol. \bibinfo{volume}{632}
  (\bibinfo{publisher}{Springer}, \bibinfo{year}{2009}).

\bibitem[{\citenamefont{Zipf}(1949)}]{Zipf49}
\bibinfo{author}{\bibfnamefont{G.}~\bibnamefont{Zipf}},
  \emph{\bibinfo{title}{Human behavior and the principle of least effort}}
  (\bibinfo{publisher}{Addison-Wesley Press, Cambridge, Mass., USA},
  \bibinfo{year}{1949}).

\bibitem[{\citenamefont{Axtell}(2001)}]{Axtell01}
\bibinfo{author}{\bibfnamefont{R.}~\bibnamefont{Axtell}},
  \bibinfo{journal}{Science} \textbf{\bibinfo{volume}{293}},
  \bibinfo{pages}{1818} (\bibinfo{year}{2001}).

\bibitem[{\citenamefont{Gabaix}(1999)}]{Gabaix99}
\bibinfo{author}{\bibfnamefont{X.}~\bibnamefont{Gabaix}},
  \bibinfo{journal}{Quart. J. Econ.} \textbf{\bibinfo{volume}{114}},
  \bibinfo{pages}{739} (\bibinfo{year}{1999}).

\bibitem[{\citenamefont{Kong et~al.}(2008)\citenamefont{Kong, Sarshar, and
  Roychowdhury}}]{Kongetal08}
\bibinfo{author}{\bibfnamefont{J.}~\bibnamefont{Kong}},
  \bibinfo{author}{\bibfnamefont{N.}~\bibnamefont{Sarshar}}, \bibnamefont{and}
  \bibinfo{author}{\bibfnamefont{V.}~\bibnamefont{Roychowdhury}},
  \bibinfo{journal}{Proc. Natl. Acad. Sci. USA} \textbf{\bibinfo{volume}{105}},
  \bibinfo{pages}{13724} (\bibinfo{year}{2008}).

\bibitem[{\citenamefont{Maillart et~al.}(2008)\citenamefont{Maillart, Sornette,
  Spaeth, and von Krogh}}]{tmaillart}
\bibinfo{author}{\bibfnamefont{T.}~\bibnamefont{Maillart}},
  \bibinfo{author}{\bibfnamefont{D.}~\bibnamefont{Sornette}},
  \bibinfo{author}{\bibfnamefont{S.}~\bibnamefont{Spaeth}}, \bibnamefont{and}
  \bibinfo{author}{\bibfnamefont{G.}~\bibnamefont{von Krogh}},
  \bibinfo{journal}{Phys. Rev. Lett.} \textbf{\bibinfo{volume}{101}},
  \bibinfo{pages}{218701(1)} (\bibinfo{year}{2008}).

\bibitem[{\citenamefont{Adamic and Huberman}(2000)}]{Huberman1}
\bibinfo{author}{\bibfnamefont{L.}~\bibnamefont{Adamic}} \bibnamefont{and}
  \bibinfo{author}{\bibfnamefont{B.}~\bibnamefont{Huberman}},
  \bibinfo{journal}{Quarterly Journal of Electronic Commerce}
  \textbf{\bibinfo{volume}{1}}, \bibinfo{pages}{5} (\bibinfo{year}{2000}).

\bibitem[{\citenamefont{Furusawa and Kaneko}(2003)}]{FurusawaKaneko03}
\bibinfo{author}{\bibfnamefont{C.}~\bibnamefont{Furusawa}} \bibnamefont{and}
  \bibinfo{author}{\bibfnamefont{K.}~\bibnamefont{Kaneko}},
  \bibinfo{journal}{Phys. Rev. Lett.} \textbf{\bibinfo{volume}{90}},
  \bibinfo{pages}{088102} (\bibinfo{year}{2003}).

\bibitem[{\citenamefont{Simon}(1955)}]{Simon55}
\bibinfo{author}{\bibfnamefont{H.}~\bibnamefont{Simon}},
  \bibinfo{journal}{Biometrika} \textbf{\bibinfo{volume}{52}},
  \bibinfo{pages}{425} (\bibinfo{year}{1955}).

\bibitem[{\citenamefont{Simon}(1960)}]{Simon60}
\bibinfo{author}{\bibfnamefont{H.}~\bibnamefont{Simon}},
  \bibinfo{journal}{Information and Control} \textbf{\bibinfo{volume}{3}},
  \bibinfo{pages}{80} (\bibinfo{year}{1960}).

\bibitem[{\citenamefont{Ijri and Simon}(1977)}]{IS77}
\bibinfo{author}{\bibfnamefont{Y.}~\bibnamefont{Ijri}} \bibnamefont{and}
  \bibinfo{author}{\bibfnamefont{H.~A.} \bibnamefont{Simon}},
  \emph{\bibinfo{title}{Distributions and the Sizes of Business Firms}}
  (\bibinfo{publisher}{North-Holland, New York}, \bibinfo{year}{1977}).

\bibitem[{\citenamefont{Gibrat}(1931)}]{Gibrat31}
\bibinfo{author}{\bibfnamefont{R.}~\bibnamefont{Gibrat}},
  \emph{\bibinfo{title}{Les In\'egalit\'es Economiques}}
  (\bibinfo{publisher}{Librairie du Recueil Sirey, Paris},
  \bibinfo{year}{1931}).

\bibitem[{\citenamefont{Barabasi and Albert}(1999)}]{AlbertBaraAlbert99}
\bibinfo{author}{\bibfnamefont{A.-L.} \bibnamefont{Barabasi}} \bibnamefont{and}
  \bibinfo{author}{\bibfnamefont{R.}~\bibnamefont{Albert}},
  \bibinfo{journal}{Science} \textbf{\bibinfo{volume}{286}},
  \bibinfo{pages}{509} (\bibinfo{year}{1999}).

\bibitem[{\citenamefont{Kesten}(1973)}]{Kesten73}
\bibinfo{author}{\bibfnamefont{H.}~\bibnamefont{Kesten}},
  \bibinfo{journal}{Acta Math.} \textbf{\bibinfo{volume}{131}},
  \bibinfo{pages}{207} (\bibinfo{year}{1973}).

\bibitem[{\citenamefont{Sornette}(1998)}]{Sornette98}
\bibinfo{author}{\bibfnamefont{D.}~\bibnamefont{Sornette}},
  \bibinfo{journal}{Phys. Rev. E} \textbf{\bibinfo{volume}{57}},
  \bibinfo{pages}{4811} (\bibinfo{year}{1998}).

\bibitem[{\citenamefont{Gabaix}(2009)}]{Gabaix2009}
\bibinfo{author}{\bibfnamefont{X.}~\bibnamefont{Gabaix}},
  \bibinfo{journal}{Ann. Rev. Econ.} \textbf{\bibinfo{volume}{1}},
  \bibinfo{pages}{255} (\bibinfo{year}{2009}).

\bibitem[{\citenamefont{Malevergne et~al.}(2010)\citenamefont{Malevergne,
  Saichev, and Sornette}}]{dsornette2}
\bibinfo{author}{\bibfnamefont{Y.}~\bibnamefont{Malevergne}},
  \bibinfo{author}{\bibfnamefont{A.}~\bibnamefont{Saichev}}, \bibnamefont{and}
  \bibinfo{author}{\bibfnamefont{D.}~\bibnamefont{Sornette}},
  \bibinfo{journal}{preprint}  (\bibinfo{year}{2010}),
  \bibinfo{note}{\url{http://ssrn.com/abstract=1083962}}.

\end{thebibliography}

\end{document}